\documentclass[letterpaper]{article}
\usepackage{aaai20}  
\usepackage{times}  
\usepackage{helvet} 
\usepackage{courier}  
\usepackage[hyphens]{url}  
\usepackage{graphicx} 
\usepackage{graphicx}  
\usepackage{verbatim}
\usepackage{amsmath}
\usepackage{amssymb}
\usepackage{amsthm}
\usepackage{rotating}
\usepackage{algorithm}
\usepackage[noend]{algpseudocode}

\usepackage{tabularx}
\usepackage{subfigure}
\usepackage{multirow}
\usepackage{siunitx}

\usepackage{fancyhdr}
\fancypagestyle{firststyle}
{
   \fancyhf{}
   \fancyfoot[C]{``A'' (Approved for Public Release, Distribution Unlimited)}
}

\newtheorem{definition}{Definition}
\newtheorem{theorem}{Theorem}
\newtheorem{proposition}{Proposition}

\begin{document}
\title{Parallel Algorithm for Approximating Nash Equilibrium in Multiplayer Stochastic Games with Application to Naval Strategic Planning\thanks{This research was developed with funding from the Defense Advanced Research Projects Agency (DARPA). The views, opinions and/or findings expressed are those of the author and should not be interpreted as representing the official views or policies of the Department of Defense or the U.S. Government.}}
\author{Sam Ganzfried$^1$, Conner Laughlin$^2$, Charles Morefield$^2$\\ 
$^1$Ganzfried Research $^2$Arctan, Inc.
}

\maketitle
\thispagestyle{firststyle}

\begin{abstract}
Many real-world domains contain multiple agents behaving strategically with probabilistic transitions and uncertain (potentially infinite) duration. Such settings can be modeled as stochastic games. While algorithms have been developed for solving (i.e., computing a game-theoretic solution concept such as Nash equilibrium) two-player zero-sum stochastic games, research on algorithms for non-zero-sum and multiplayer stochastic games is limited. We present a new algorithm for these settings, which constitutes the first parallel algorithm for multiplayer stochastic games. We present experimental results on a 4-player stochastic game motivated by a naval strategic planning scenario, showing that our algorithm is able to quickly compute strategies constituting Nash equilibrium up to a very small degree of approximation error. 
\end{abstract}

\section{Introduction}
\label{se:introduction}
Nash equilibrium has emerged as the most compelling solution concept in multiagent strategic interactions. For two-player zero-sum (adversarial) games, a Nash equilibrium can be computed in polynomial time (e.g., by linear programming). This result holds both for simultaneous-move games (often represented as a matrix), and for sequential games of both perfect and imperfect information (often represented as an extensive-form game tree). However, for non-zero-sum and games with 3 or more agents it is PPAD-hard to compute a Nash equilibrium (even for the simultaneous-move case) and widely believed that no efficient algorithms exist~\cite{Chen05:Nash,Chen06:Settling,Daskalakis09:Complexity}. For simultaneous (\emph{strategic-form}) games several approaches have been developed with varying degrees of success~\cite{Berg17:Exclusion,Porter08:Simple,Govindan03:Global,Lemke64:Equilibrium}.

While extensive-form game trees can be used to model sequential actions of a known duration (e.g., repeating a simultaneous-move game for a specified number of iterations), they cannot model games of unknown duration, which can potentially contain infinite cycles between states. Such games must be modeled as \emph{stochastic games}.  

\begin{definition}
A \emph{stochastic game} (aka \emph{Markov game}) is a tuple $(Q, N, A, P, r)$, where:
\begin{itemize}
\item $Q$ is a finite set of (stage) games (aka \emph{game states})
\item $N$ is a finite set of $n$ players
\item $A = A_1 \times \ldots \times A_n$, where $A_i$ is a finite set of actions available to player $i$
\item $P : Q \times A \times Q \rightarrow [0,1]$ is the transition probability function; $P(q,a,\hat{q})$ is the probability of transitioning from state $q$ to state $\hat{q}$ after action profile $a$
\item $R = r_1,\ldots,r_n$, where $r_i : Q \times A \rightarrow \mathbb{R}$ is a real-valued payoff function for player $i$
\end{itemize}
\end{definition}

There are two commonly used methods for aggregating the stage game payoffs into an overall payoff: \emph{average (undiscounted) reward} and \emph{future discount reward} using a discount factor $\delta < 1$. Stochastic games generalize several commonly-studied settings, including games with finite interactions, strategic-form games, repeated games, stopping games, and Markov decision problems.

The main solution concept for stochastic games, as for other game classes, is Nash equilibrium (i.e., a strategy profile for all players such that no player can profit by unilaterally deviating), though some works have considered alternative solution concepts such as correlated equilibrium and Stackelberg equilibrium. Before discussing algorithms, we point out that, unlike other classes of games such as strategic and extensive-form, it is not guaranteed that Nash equilibrium exists in general in stochastic games. 

One theorem states that if there is a finite number of players and the action sets and the set of states are finite, then a stochastic game with a finite number of stages always has a Nash equilibrium (using both average and discounted reward). Another result shows that this is true for a game with infinitely many stages if total payoff is the discounted sum.

Often a subset of the full set of strategies is singled out called \emph{stationary strategies}. A strategy is \emph{stationary} if it depends only on the current state (and not on the time step). Note that in general a strategy could play different strategies at the same game state at different time steps, and a restriction to stationary strategies results in a massive reduction in the size of the strategy spaces to consider. It has been shown that in two-player discounted stochastic games there exists an equilibrium in stationary policies. For the undiscounted (average-reward) setting, it has recently been proven that each player has a strategy that is $\epsilon$-optimal in the limit as $\epsilon \rightarrow 0$, technically called a \emph{uniform equilibrium}, first for two-player zero-sum games~\cite{Mertens81:Stochastic} and more recently for general-sum games~\cite{Vieille00:Two-player}. 

Thus, overall, the prior results show that for two-player (zero-sum and non-zero-sum) games there exists an equilibrium in stationary strategies for the discounted reward model, and a uniform equilibrium for the average reward model. However, for more than two players, only the first of these is guaranteed, and it remains an open problem whether a (uniform) equilibrium exists in the undiscounted average-reward model. Perhaps this partially explains the scarcity of research on algorithms for multiplayer stochastic games.

Several stochastic game models have been proposed for national security settings. For example, two-player discounted models of adversarial patrolling have been considered, for which mixed-integer program formulations are solved to compute a Markov stationary Stackelberg equilibrium~\cite{Vorobeychik12:Computing,Vorobeychik14:Computing}. One work has applied an approach to approximate a correlated equilibrium in a three-player threat prediction game model~\cite{Chen07:Game}. However we are not aware of other prior research on settings with more than two players with guarantees on solution quality (or for computing Nash as opposed to Stackelberg or correlated equilibrium).

The only prior research we are aware of for computing Nash equilibria in multiplayer stochastic games has been approaches developed for poker tournaments~\cite{Ganzfried08:Computing,Ganzfried09:Computing}. Our algorithms are based on the approaches developed in that work. The model was a 3-player poker tournament, where each game state corresponded to a vector of stack sizes. The game had potentially infinite duration (e.g., if all players continue to fold the game could continue arbitrarily long), and was modeled assuming no discount factor. Several algorithms were provided, with the best-performer being based on integrating fictitious play (FP) with a variant of policy iteration. While the algorithm is not guaranteed to converge, a technique was developed that computes the maximum amount a player could gain by deviating from the computed strategies, and it was verified that in fact this value was quite low, demonstrating that the algorithm successfully computed a very close approximation of Nash equilibrium. In addition to being multiplayer, this model also differed from the previous models in that the stage games had imperfect information.

The main approaches from prior work on multiplayer stochastic game solving integrate algorithms for solving stage games (of imperfect information) assuming specified values for the payoffs of all players at transitions into other stage games, and techniques for updating the values for all players at all states in light of these newly computed strategies. For the stage game equilibrium computation these algorithms used fictitious play, which is an iterative algorithm that has been proven to converge to Nash equilibrium in certain classes of games (two-player zero-sum and certain classes of two-player general-sum). For multiplayer and non-zero-sum games it does not guarantee convergence to equilibrium, and all that can be proven is that if it does happen to converge then the sequence of strategies determined by the iterations constitutes an equilibrium. It did happen to converge consistently in the 3-player application despite the fact that it is not guaranteed to do so, suggesting that it likely performs better in practice than the worst-case theory would dictate. For the value updating step, variants of value iteration and policy iteration (which are approaches for solving Markov decision processes) were used.   

Note that there has been significant recent attention on an alternative iterative self-play algorithm known as counterfactual regret minimization (CFR). Like FP, CFR is proven to converge to a Nash equilibrium in the limit for two-player zero-sum games. 
For multiplayer and non-zero-sum games the algorithm can also be run, though the strategies computed are not guaranteed to form a Nash equilibrium. It was demonstrated that it does in fact converge to an $\epsilon$-Nash equilibrium (a strategy profile in which no agent can gain more than $\epsilon$ by deviating) in the small game of three-player Kuhn poker, while it does not converge to equilibrium in Leduc hold 'em~\cite{Abou10:Using}. It was subsequently proven that it guarantees converging to a strategy that is not dominated and does not put any weight on iteratively weakly-dominated actions~\cite{Gibson14:Regret}. While for some small games this guarantee can be very useful (e.g., for two-player Kuhn poker a high fraction of the actions are iteratively-weakly-dominated), in many large games (such as full Texas hold 'em) only a very small fraction of actions are dominated, and the guarantee is not useful~\cite{Ganzfried19:Mistakes}. Very recently an agent based on CFR has defeated strong human players in a multiplayer poker cash game\footnote{Note that a poker cash game is modeled as a standard extensive-form game, while the poker tournament described above is modeled as a stochastic game. In a cash game chips represent actual money, while in a tournament chips have no monetary value and are only a proxy, as players receive money only after they run out of chips (depending on their position of elimination).}~\cite{Brown19:Superhuman}. However, much of the strength of the agent came from real-time solving of smaller portions of the game which typically contained just two players using ``endgame''/``subgame'' solving~\cite{Ganzfried15:Endgame} and more recently depth limited ``midgame'' solving~\cite{Hu17:Midgame,Brown18:Depth}. 
Recently it has been shown that when integrated with deep learning a version of CFR outperforms FP in two-player zero-sum poker variants~\cite{Brown19:Deep}, though the core version of FP outperforms CFR in multiplayer and non-zero-sum settings~\cite{Ganzfried20:Fictitious}.

In this work we build on the prior algorithms for multiplayer stochastic games to solve a 4-player model of naval strategic planning that we refer to as a \emph{Hostility Game}. This is a novel model of national security that has been devised by a domain expert. The game is motivated by the Freedom of Navigation Scenario in the South China Sea, though we think it is likely also applicable to other situations, and in general that multiplayer stochastic games are fundamental for modeling national security scenarios. 

\section{Hostility game}
\label{se:hostility}
\begin{figure}[!ht]
\centering
\includegraphics[scale=0.2]{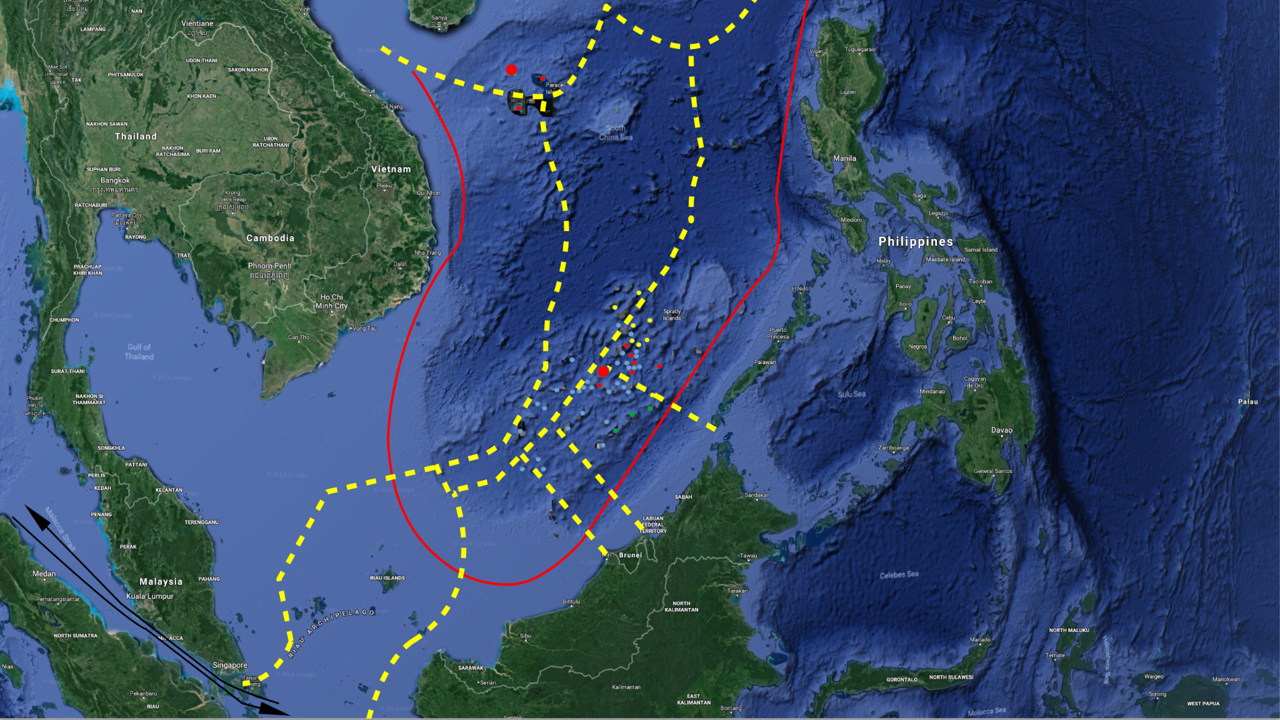}
\caption{General figure for South China Sea scenario.}
\label{fi:SCS}
\end{figure}

In the South China Sea a set of \emph{blue} players attempts to navigate freely, while a set of \emph{red} players attempt to obstruct this from occurring (Figure~\ref{fi:SCS}). 
In our model there is a single blue player and several red players of different ``types'' which may have different capabilities (we will specifically focus on the setting where there are three different types of red players). If a blue player and a subset of the red players happen to navigate to the same location, then a confrontation will ensue, which we refer to as a Hostility Game.

\begin{figure}[!ht]
\centering
\includegraphics[scale=0.4]{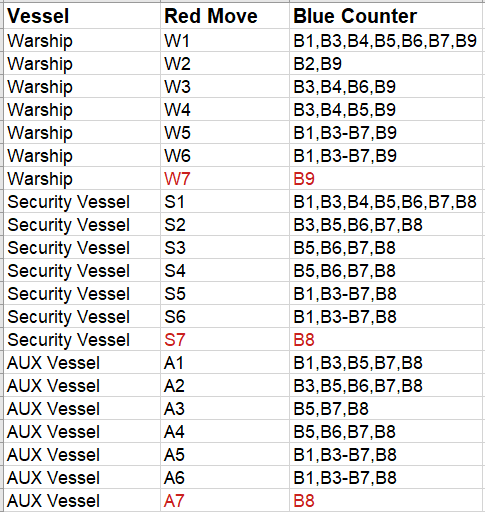}
\caption{List of blue moves that counter each red move.}
\label{fi:countered}
\end{figure}

\begin{figure*}[!ht]
\centering
\includegraphics[scale=0.2]{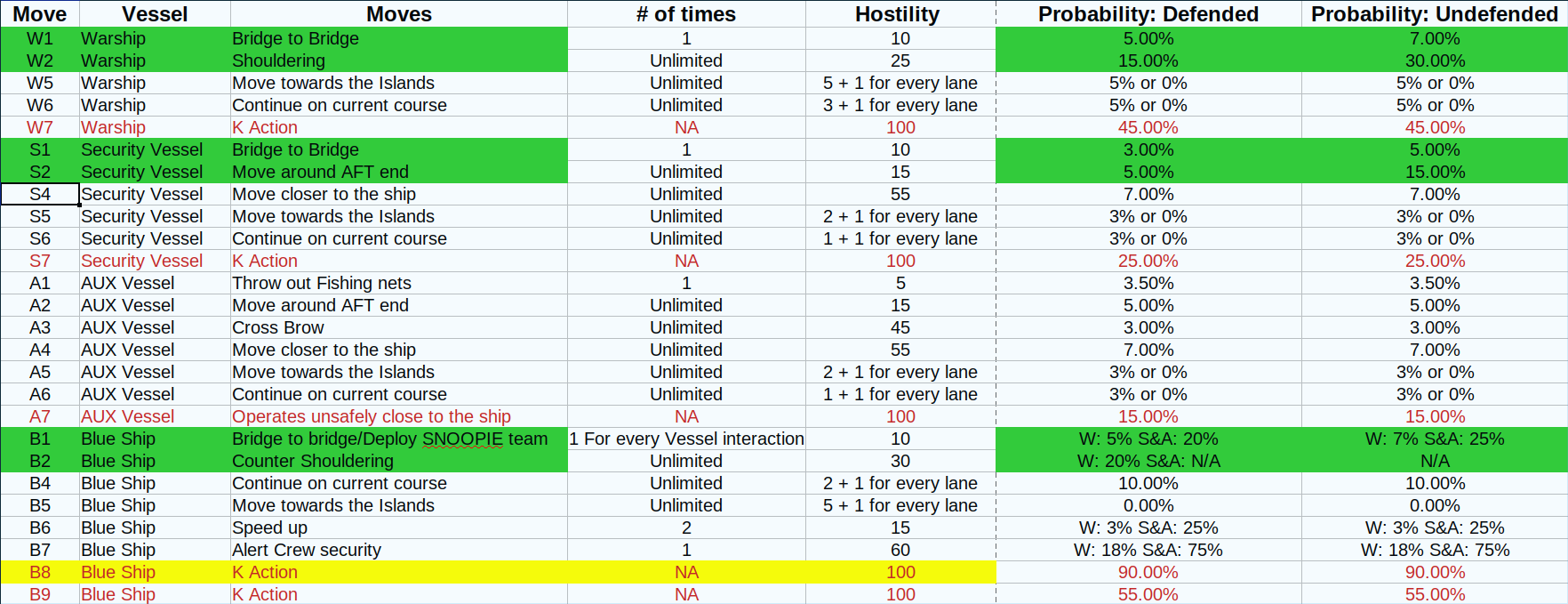}
\caption{Sample of typical actions and parameters for Hostility Game.}
\label{fi:hostility}
\end{figure*}

In a Hostility Game, each player can initially select from a number of available actions (which is between 7 and 10 for each player). Certain actions for the blue player are \emph{countered} by certain actions of each of the red players, while others are not (Figure~\ref{fi:countered}). Depending on whether the selected actions constitute a counter, there is some probability that the blue player \emph{wins} the confrontation, some probability that the red players win, and some probability that the game repeats. Furthermore, each action of each player has an associated \emph{hostility level}. Initially the game starts in a state of zero hostility, and if it is repeated then the overall hostility level increases by the sum of the hostilities of the selected actions. If the overall hostility level reaches a certain threshold (300), then the game goes into \emph{kinetic mode} and all players achieve a very low payoff (negative 200). If the game ends in a win for the blue player, then the blue player receives a payoff of 100 and the red players receive negative 100 (and vice versa for a red win). Note that the game repeats until either the blue/red players win or the game enters kinetic mode. A subset of the game's actions and parameters are given in Figure~\ref{fi:hostility}. Note that in our model we assume that all red players act independently and do not coordinate their actions. Our game model and parameters were constructed from discussions with a domain expert.

\begin{definition}
A hostility game (HG) is a tuple \\ $G = (N, M, c, b^D, b^U, r^D, r^U, \pi, h, K, \pi^K)$, where
\begin{itemize}
\item $N$ is the set of players. For our initial model we will assume player 1 is a blue player and players 2--4 are red players (P2 is a Warship, P3 is a Security ship, and P4 is an Auxiliary vessel). 
\item $M = \{M_i\}$ is the set of actions, or \emph{moves}, where $M_i$ is the set of moves available to player $i$
\item For $m_i \in M_i$, $c(M_i)$ gives a set of blue moves that are \emph{counter moves} of $m_i$
\item For each blue player move and red player, a probability of blue success/red failure given that the move is defended against (i.e., \emph{countered}), denoted as $b^D$
\item Probability that a move is a blue success/red failure given the move is Undefended against, denoted as $b^U$
\item Probability for a red success/blue failure given the move is defended against, $r^D$
\item Probability for a red success/blue failure given the move is undefended against, $r^U$
\item Real valued payoff for success for each player, $\pi_i$
\item Real-valued hostility level for each move $h(m_i)$
\item Positive real-valued kinetic hostility threshold $K$
\item Real-valued payoffs for each player when game goes into Kinetic mode, $\pi^K_i$
\end{itemize}
\end{definition}

We model hostility game $G$ as a (4-player) stochastic game with a collection of stage games $\{G_n\},$ where $n$ corresponds to the cumulative sum of hostility levels of actions played so far. The game has $K+3$ states: $G_0,\ldots,G_K$, with two additional terminal states B and R for blue and red victories. Depending on whether the blue move is countered, there is a probabilistic outcome for whether the blue player or red player (or neither) will outright win. The game will then transition into terminal states B or R with these probabilities, and then will be over with final payoffs. Otherwise, the game transitions into $G_{n'}$ where $n'$ is the new sum of the hostility levels. If the game reaches $G_K$, the players obtain the kinetic payoff $\pi^K_i$. Thus, the game starts at initial state $G_0$ and after a finite number of time steps will eventually reach one of the three terminal states $B, R,$ or $G_K$.

Note that in our formulation there is a finite number of players (4) as well as a finite number of states ($K+3$). Furthermore, with the assumption that hostility levels for all actions are positive, the game must complete within a finite number of stages (because the combined hostility level will ultimately reach $K$ if one of the terminal states $B$ or $R$ is not reached before then). So a Nash equilibrium is guaranteed to exist in stationary strategies, for both the average and discounted reward models. 
Note that the payoffs are only obtained in the final stage when a terminal state is reached, and so the difference between using average and discounted reward is likely less significant than for games where rewards are frequently accumulated within different time steps. 



\section{Algorithm}
\label{se:algorithm}
While research on algorithms for stochastic games with more than two players is limited, several prior algorithms have been devised and applied in the context of a poker tournament~\cite{Ganzfried08:Computing,Ganzfried09:Computing}. At a high level these algorithms consist of two different components: first is a \emph{game-solving algorithm} that computes an (approximate) Nash equilibrium at each stage game assuming given values for all players at the other states, and the second is a \emph{value update} procedure that updates values for all players at all states in light of the newly-computed stage-game strategies. For the poker application the stage games were themselves games of imperfect information (the players must select a strategy for every possible set of private cards that they could hold at the given vector of chip stack sizes). The fictitious play algorithm was used for the game-solving step, which applies both to games of perfect and imperfect information. Fictitious play is an iterative self-play algorithm that has been proven to converge to Nash equilibrium in certain classes of games (two-player zero-sum and certain non-zero-sum). For multiplayer and non-zero-sum games it does not guarantee convergence to equilibrium, and all that can be proven is that if it does happen to converge, then the sequence of strategies determined by the iterations constitutes an equilibrium (Theorem~\ref{th:fp}). It did happen to converge consistently in the 3-player application despite the fact that it is not guaranteed to do so, suggesting that it likely performs better in practice than the worst-case theory would dictate.

In (smoothed) fictitious play each player $i$ plays a best response to the average opponents' strategies thus far, using the following rule at time $t$ to obtain the current strategy,
$$s^t _i = \left( 1 - \frac{1}{t} \right) s^{t-1} _i + \frac{1}{t} s'^t _i,$$
where $s'^t _i$ is a best response of player $i$ to the profile $s^{t-1} _{-i}$  of the other players played at time $t-1$ (strategies can be initialized arbitrarily at $t=0$, and for our experiments we will initialize them to be uniformly random).  This algorithm was originally developed as a simple learning model for repeated games, and was proven to converge to a Nash equilibrium in two-player zero-sum games~\cite{Fudenberg98:Theory}.  However, it is not guaranteed to converge in two-player general-sum games or games with more than two players. All that is known is that if it does converge, then the strategies constitute a Nash equilibrium (Theorem~\ref{th:fp}).

\begin{theorem}
~\cite{Fudenberg98:Theory} Under fictitious play, if the empirical distributions over each player's choices converge, the strategy profile corresponding to the product of these distributions is a Nash equilibrium.
\label{th:fp}
\end{theorem}

A meta-algorithm that integrates these two components---stage game solving and value updating---is depicted in Algorithm~\ref{al:meta}. We initialize the values at all states according to $V_0$, and alternate between the phase of solving each nonterminal stage game using algorithm $A$ (note that for certain applications it may even make sense to use a different stage game algorithm $A_i$ for different states), and the value update phase using algorithm $V$. Following prior work we will be using fictitious play for $A$ and variants of value and policy iteration for $V$, though the meta-algorithm is general enough to allow for alternative choices depending on the setting.

\begin{algorithm}
\caption{Meta-algorithm for multiplayer stochastic game equilibrium computation}
\label{al:meta}
\textbf{Inputs}: Stochastic game $G$ with set of terminal states $\{T_n\}$ and set of $U$ nonterminal states $\{U_n\}$, algorithm for stage game equilibrium computation $A$, algorithm
for updating values of all nonterminal states for all players $V$, number of iterations $N$, initial assignment of state values $V_0$.
\begin{algorithmic}
\State Initialize values for all players for all nonterminal states according to $V_0$.
\For {$n = 1$ to $N$}
\For {$i = 1$ to $U$}
\State Solve stage game defined at $U_i$ using algorithm $A$ assuming values given by $V_{n-1}$. 
\State Let $S_{i,n}$ denote the equilibrium for state $i$.
\EndFor 
\State Update the values for all nonterminal states $U_i$ according to algorithm $V$ assuming that strategies $S_{i,n}$ are used at game state $U_i$. 
\EndFor
\State Output strategies $\{S_{i,N}\}$
\end{algorithmic}
\end{algorithm}

The first algorithm previously considered, called VI-FP, instantiates Algorithm~\ref{al:meta} using fictitious play for solving stage games and a multiplayer analogue of \emph{value iteration} for updating values~\cite{Ganzfried08:Computing,Ganzfried09:Computing}. As originally implemented (Algorithm~\ref{al:VIFP}), the algorithm takes two inputs, which determine the stopping criterion for the two phases. The fictitious play phase halts on a given state when no player can gain more than $\gamma$ by deviating from the strategies (i.e., the strategies constitute a $\gamma$-equilibrium), and the value iteration phase halts when all game state values for all players change by less than $\delta$.   

\begin{algorithm}
\caption{VI-FP~\cite{Ganzfried09:Computing}}
\label{al:VIFP}
\textbf{Inputs}: Degree of desired stage game solution approximation $\gamma$, desired
max difference between value updates $\delta$   
\begin{algorithmic}
\State $V^0 =$ initializeValues()
\State diff $= \infty$
\State $i = 0$
\While {diff $> \delta$}
\State $i = i+1$
\State regret = $\infty$
\State $S$ = initializeStrategies()
\While {regret $> \gamma$}
\State $S$ = fictPlay()
\State regret = maxRegret($S$)
\EndWhile
\State $V^i = $ getNewValues($V^{i-1}$,$S$)
\State diff = maxDev($V^i, V^{i-1})$
\EndWhile
\Return{$S$}
\end{algorithmic}
\end{algorithm}

Prior work used a domain-specific initialization for the values $V^0$ called the Independent Chip Model for poker tournaments~\cite{Ganzfried08:Computing}. A counterexample was provided showing that VI-FP may actually converge to non-equilibrium strategies if a poor initialization is used~\cite{Ganzfried09:Computing}, and it was suggested based on a prior theorem for value iteration in single-agent Markov decision processes (MDPs) that this phenomenon can only occur if not all values are initialized pessimistically (Theorem~\ref{th:value-iteration}). We note that there is not a well-defined notion of $v^*$ in our setting, as multiplayer games can contain multiple Nash equilibria yielding different payoffs to the players. 

\begin{theorem}
~\cite{Puterman05:Markov} In our setting, if $v^0$ is initialized pessimistically (i.e., $\forall s, v^0(s) \leq v^*(s)$), value iteration converges (pointwise and monotonically) to $v^*.$
\label{th:value-iteration}
\end{theorem}

We also note that the prior work proposed just one option for a set of halting criteria for fictitious play and value iteration. Since fictitious play is not guaranteed to converge in multiplayer games there is no guarantee that the approximation threshold of $\gamma$ will be reached for sufficiently small values (and similarly there is no guarantee that a value difference threshold of $\delta$ will be obtained for the outer loop). There are several other sensible choices of halting criteria, for example running the algorithms for a specified number of iterations as we have done in our meta-algorithm, Algorithm~\ref{al:meta}. As we will see when we describe our parallel algorithm, this approach would also allow for more consistency between the runtimes of computations on different cores. Another halting criterion for fictitious play is to run it for a specified number of iterations but output the average strategies that produced lowest approximation error $\epsilon$ out of all iterations (not just the final strategies after the last iteration).

The next approach considered by prior work also used fictitious play for the stage-game solving phase but substituted in a variant of the policy-iteration algorithm (Algorithm~\ref{al:policy-iteration}) for value iteration in the value update phase. This algorithm called PI-FP is depicted in Algorithm~\ref{al:PIFP}. The new values are computed by solving a system of equations defined by a transition matrix. In effect this corresponds to updating all game state values globally to be consistent with the recently-computed stage game strategies, while the value iteration procedure updates the values locally given the prior values of the adjacent states. Thus, at least intuitively we would likely expect PI-FP to outperform VI-FP for this reason. Unlike VI-FP, for PI-FP it can be proven (Proposition~\ref{pr:PIFP}) that if the algorithm converges then the resulting strategies constitute a Nash equilibrium (regardless of the initialization). The experimental results of prior work agreed with this intuition, as PI-FP converged to near-equilibrium faster than VI-FP~\cite{Ganzfried09:Computing}. This was determined by an ex-post checking procedure to compute the degree of approximation $\epsilon$ given by Algorithm~\ref{al:epc}, with correctness following from Theorem~\ref{th:policy-iteration} for Algorithm~\ref{al:policy-iteration}. The quantity $v_i^{\pi^*_i,s^*_{-i}}(G_0)$ denotes the value to player $i$ at the initial game state when player $i$ plays $\pi^*_i$ and his opponents play $s^*_{-i}$, and $v_i^{s^*_i,s^*_{-i}}(G_0)$ is analogous.

\begin{algorithm}
\caption{PI-FP~\cite{Ganzfried09:Computing}}
\textbf{Inputs}: Degree of desired stage game solution approximation $\gamma$, desired
max difference between value updates $\delta$   
\label{al:PIFP}
\begin{algorithmic}
\State $V^0 =$ initializeValues()
\State diff $= \infty$
\State $i = 0$
\While {diff $> \delta$}
\State $i = i+1$
\State regret = $\infty$
\State $S^0$ = initializeStrategies()
\While {regret $> \gamma$}
\State $S^i$ = fictPlay()
\State regret = maxRegret($S^i$)
\EndWhile
\State $M^i = $ createTransitionMatrix($S^i$)
\State $V^i = $ evaluatePolicy($M^i$)
\State diff = maxDev($V^i, V^{i-1})$
\EndWhile
\Return{$S^i$}
\end{algorithmic}
\end{algorithm}

\begin{proposition}
If the sequence of strategies $\{s^n\}$ determined by iterations of the outer loop of Algorithm~\ref{al:PIFP} converges, then the final strategy profile $s^*$ is an equilibrium.
\label{pr:PIFP}
\end{proposition}

\begin{algorithm}
\caption{Policy iteration for positive bounded models with expected total-reward criterion}
\label{al:policy-iteration}
\begin{enumerate}
\item Set $n = 0$ and initialize the policy $\pi^0$ so it has nonnegative expected reward.

\item Let $v^n$ be the solution to the system of equations
$$v(i) = r(i) + \sum_j p_{ij}^{\pi^n} v(j)$$
where $p_{ij}^{\pi^n}$ is the probability of moving from state $i$ to state $j$ under policy $\pi^n$.  If there are multiple solutions, let $v^n$ be the minimal nonnegative solution.

\item For each state $s$ with action space $A(s)$, set
$$\pi^{n+1}(s) \in \mathrm{arg}\hspace{-0.1cm}\max_{\hspace{-0.3cm} a \in A(s)} \sum_j p_{ij}^a v^n(j),$$
breaking ties so $\pi^{n+1}(s) = \pi^n(s)$ whenever possible.

\item If $\pi^{n+1}(s) = \pi^n(s)$ for all $s$, stop and set $\pi^* = \pi^n.$  Otherwise increment $n$ by 1 and return to Step~2.
\end{enumerate}
\end{algorithm}

\begin{theorem}
\label{th:policy-iteration}
~\cite{Puterman05:Markov} Let $S$ be the set of states in $M$.  Suppose $S$ and $A(s)$ are finite. Let $\{v^n\}$ denote the sequence of iterates of Algorithm~\ref{al:policy-iteration}.
Then, for some finite $N$, $v^N = v^*$ and $\pi^N = \pi^*$.
\end{theorem}

\begin{algorithm}
\caption{\emph{Ex post} check procedure}
\label{al:epc}
\small 
\begin{algorithmic}
\State Create MDP $M$ from the strategy profile $s^*$
\State Run Algorithm~\ref{al:policy-iteration} on $M$ (using initial policy $\pi^0 = s^*$) to get $\pi^*$
\State \Return {$\max_{i \in N} \left[ v_i^{\pi^*_i,s^*_{-i}}(G_0) - v_i^{s^*_i,s^*_{-i}}(G_0) \right]$}
\end{algorithmic}
\end{algorithm}
\normalsize

\begin{proposition}
Algorithm~\ref{al:epc} correctly computes the largest amount any agent can improve its expected utility by deviating from $s^*$.
\label{ExPostProp}
\end{proposition}

The implementations of VI-FP and PI-FP in prior work both used a single core, and involved running fictitious play sequentially at every game state within the stage game update phase. We observe that both of these approaches can be parallelized. Assuming there are $|S|$ states and $d$ cores (and for presentation simplicity assuming that $|S|$ is a multiple of $d$), we can assign $\frac{|S|}{d}$ of the stage games to each core and run fictitious play independently on $d$ states simultaneously. This will compute equilibrium strategies at all stage games, which can be integrated with the value update phase of both VI-FP and PI-FP. Since the stage game solving phase is the bottleneck step of both algorithms, this parallel algorithm will achieve an approximately linear improvement in runtime by a factor of $d$. In addition to incorporating parallelization, our Algorithm~\ref{al:Parallel-PIFP} differs from the prior approach by allowing for custom stopping conditions for the two phases. 

\begin{algorithm}
\caption{Parallel PI-FP}
\textbf{Inputs}: Stopping condition $C_S$ for stage game solving, stopping condition $C_V$ for value updating, number of cores $d$
\label{al:Parallel-PIFP}
\begin{algorithmic}
\State $V^0 =$ initializeValues()
\State $i = 0$
\While {$C_V$ not met} 
\State $i = i+1$
\While {$C_S$ not met for each stage game}
\State Run fictitious play on each stage game on $d$ cores (solving $d$ stage games simultaneously) to obtain $S^i$
\EndWhile
\State $M^i = $ createTransitionMatrix($S^i$)
\State $V^i = $ evaluatePolicy($M^i$)
\EndWhile
\Return{$S^i$}
\end{algorithmic}
\end{algorithm}

We note that neither VI-FP or PI-FP is guaranteed to converge in this setting (though it has been proven that if PI-FP converges then the resulting strategies constitute a Nash equilibrium~\cite{Ganzfried09:Computing}). 
Note that our Hostility Game does not technically fall into the positive bounded model~\cite{Puterman05:Markov}, as certain actions can obtain negative payoff. However, the main difference between policy iteration for this model (Algorithm~\ref{al:policy-iteration}) as opposed to the discounted reward model is using the minimal nonnegative solution for Step 2~\cite{Puterman05:Markov}; however, for all our experiments the transition matrix had full rank and there was a unique solution. Furthermore, in a Hostility Game the rewards are only obtained at a terminal state, and the total expected reward is clearly bounded (both in the positive and negative directions). So we can still apply these versions of value and policy iteration to (hopefully) obtain optimal solutions. Note also that for the case where all hostility levels are positive we can guarantee the game will complete within a finite duration and can apply backwards induction; but this will not work in general for the case of zero or negative hostilities where the game has potentially infinite duration, and the stochastic game-solving algorithms will be needed.

\section{Experiments}
\label{se:experiments}
\begin{figure*}[!ht]
\centering
\includegraphics[scale=0.49]{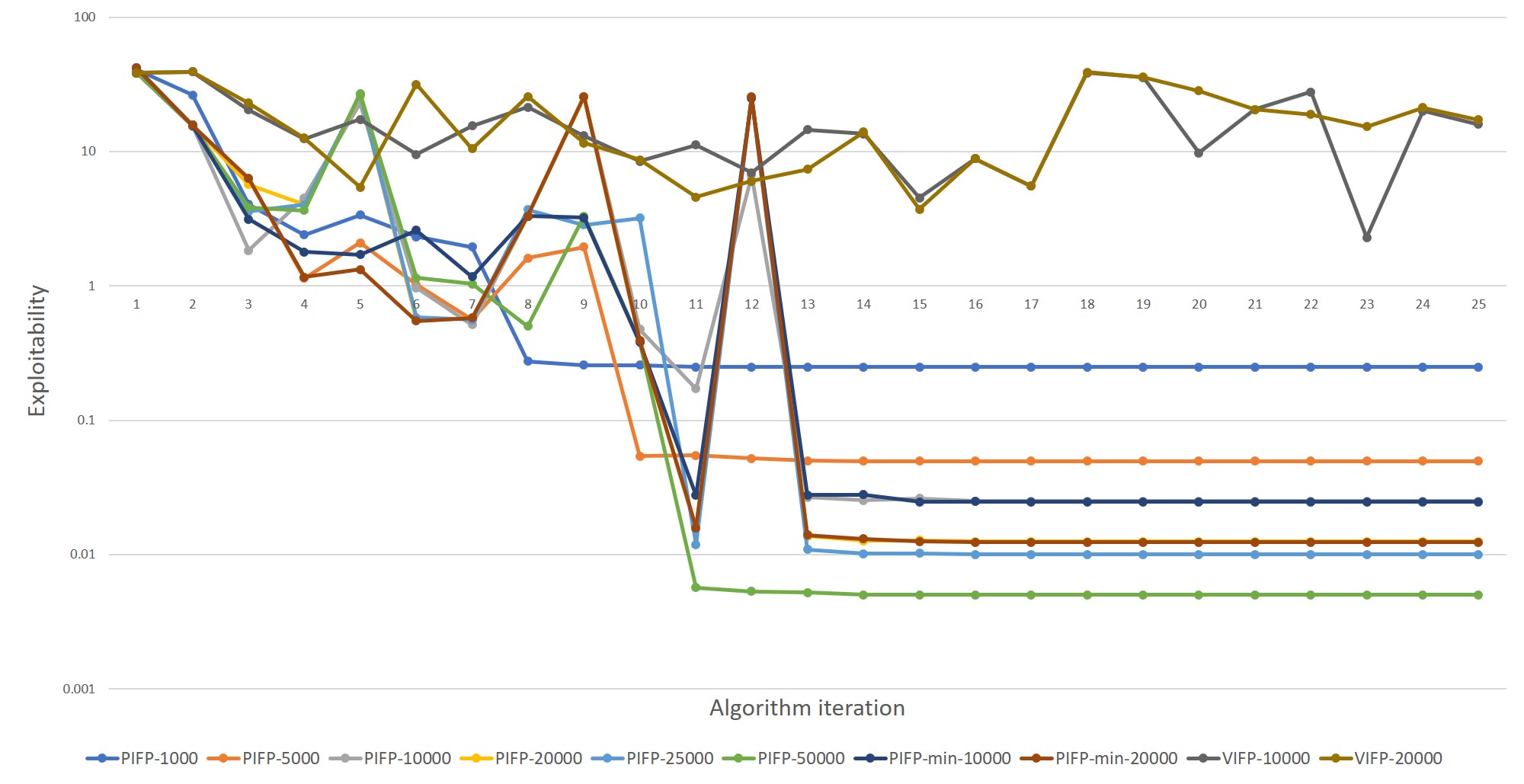}
\caption{Performance of several algorithm variants.}
\label{fi:results}
\end{figure*}

Results for the first 25 iterations of several algorithm variations are given in Figure~\ref{fi:results}. All experiments ran the parallel versions of the algorithms with 6 cores on a laptop. The variations include VI-FP and PI-FP with varying numbers of iterations of fictitious play, as well as PI-FP using the version of fictitious play where the strategy with lowest exploitability over all iterations was output (as opposed to the final strategy). We first observe that VI-FP did not converge to equilibrium while all versions of PI-FP did, making PI-FP the clearly preferable choice. We also observe that using minimum exploitability FP led to nearly identical performance as the standard version; since this version also takes longer due to the overhead of having to compute the value of $\epsilon$ at every iteration instead of just at the end, we conclude that the standard version of fictitious play is preferable to the version that selects the iteration with minimal exploitability.

For Parallel PI-FP using standard fictitious play, we compared results using 1,000, 5,000, 10,000, 20,000, 25,000, and 50,000 iterations of fictitious play for solving each game state within the inner loop of the algorithm. Each of these versions eventually converged to strategies with relatively low exploitability, with the convergence value of $\epsilon$ smaller as more iterations of FP are used. Note that initially we set values for all players at all non-terminal states to be zero, and that the terminal payoffs for a victory/loss are 100/-100, and for kinetic payoffs are -200 (with $K$=300); so convergence to $\epsilon = 0.01$ is quite good (this represents 0.01\% of the minimum possible payoff of the game). Even just using 1,000 iterations of FP converged to $\epsilon$ of around 0.25, which is still relatively small. Note that while the final convergence values were quite low, there was quite a bit of variance in $\epsilon$ for the first several iterations, even for the versions with large number of FP iterations (e.g., using 10,000--50,000 iterations spiked up to $\epsilon$ exceeding 20 at iteration 6, and using 20,000 and 25,000 spiked up again to $\epsilon$ exceeding 25 again at iteration 13). So it is very important to ensure that the algorithm can be run long enough to obtain convergence. 

\section{Conclusion}
\label{se:conclusion}
We have presented a new parallel algorithm for solving multiplayer stochastic games, and presented experimental results showing that it is able to successfully compute an $\epsilon$-equilibrium for very small $\epsilon$ for a naval strategic planning scenario that has been devised by a domain expert. 

There are several immediate avenues for future study. First, we note that while for the game model we have experimented on the stage games have perfect information, our algorithm also applies to games where the stage games have imperfect information (related prior work has shown successful convergence in the imperfect-information setting for poker tournaments). There are several different natural ways in which imperfect information can be integrated into the model. Currently we are exploring a model in which there is an unknown number of red ``sub-players'' of each of the three types; this value is known to a single ``meta-player'' of that type, but the other players only know a publicly-available distribution from which this value is drawn (much like in poker how players receive private cards known only to them and a distribution for the cards of the opponents).

We would also like to explore alternative approaches for the stage game equilibrium-computation portion of our algorithm. Currently we have used fictitious play, which has been demonstrated to obtain high performance previously. However, it may be outperformed by more recently-devised approaches such as counterfactual regret minimization. While the core version of FP has been shown to outperform CFR in multiplayer games~\cite{Ganzfried20:Fictitious}, for larger domains with complex information structures CFR may outperform fictitious play by better capitalizing on integration with forms of Monte Carlo sampling and deep learning. 

While we considered a single value for the main game parameters (set of actions, payoffs, hostility levels, etc.) that were selected by a domain expert, in practice we may not be sure of such values, and we would like to compute strategies that are robust in case our game model is inaccurate. One approach to achieve this would be to use a Bayesian setting, where the game parameters are selected according to a specified probability distribution (typically over a small number of possible options). This would require us to extend our algorithm to solve multiplayer stochastic games where the stage games are themselves Bayesian games.

While our model has assumed that the red players act independently and do not coordinate amongst themselves, this may not be the case in all realistic situations. In the extreme case when the red players are all controlled by one single meta-player, the game could simply be modeled as a two-player game (which would be zero sum for the parameters we have been using), which would be significantly easier to solve as two-player zero-sum games can be solved in polynomial time while solving multiplayer games is PPAD-hard.  We see no reason that our algorithm cannot be applied to solve alternative modifications of the model that integrate more subtle forms of coordination between players.

Our game model assumed that all hostility levels are positive, from which we are able to conclude the existence of a Nash equilibrium in stationary strategies (because the game would be guaranteed to have a finite number of stages); however, we could not make the same deduction if some hostility levels are non-positive for the undiscounted setting (though we still could if we were using discounted reward). In the future we would like to explore convergence of our algorithm for different selections of the hostility levels including zero and negative values, as well as consider potential differences between the average and discounted reward settings.

By now we have observed fictitious play to converge consistently for stage games in several domains (previously for poker tournaments and now for naval planning), as well as the general PI-FP algorithm converge for multiplayer stochastic games. Theoretically we have seen that these approaches are not guaranteed to converge in general for these game classes, and all that has been proven is that if they do converge then the computed strategies constitute a Nash equilibrium (though for VI-FP this is not the case and a counterexample was shown where it can converge to non-equilibrium strategies~\cite{Ganzfried09:Computing}). It would be interesting from a theoretical perspective to prove more general conditions for which these algorithms are guaranteed to converge in multiplayer settings that can include generalizations of these settings that have been studied.



Many important real-world settings contain multiple players interacting over an unknown duration with probabilistic transitions, and we feel that the multiplayer stochastic game model is fundamental for many national security domains, particularly with the ability of approaches to be integrated with imperfect information and Bayesian parameter uncertainty. We plan to explore the application of our algorithm to other similarly complicated domains in the near future. 

\bibliographystyle{aaai}
\bibliography{C://FromBackup/Research/refs/dairefs}

\end{document}